\documentclass[pre,twocolumn,superscriptaddress]{revtex4}
\usepackage{latexsym}
\usepackage{amssymb}
\usepackage{amsmath}

\usepackage{mathptmx}       
\usepackage{helvet}         
\usepackage{courier}        
\usepackage{amsthm,amsmath,amsfonts,amssymb,latexsym}

\newcommand{\R}{\mathbb R}

\def\be#1\ee{\begin{equation}#1\end{equation}}
\newcommand{\fer}[1]{(\ref{#1})}

\newcommand{\bq}{\begin{equation}}
\newcommand{\eq}{\end{equation}}

\def\bqa{\begin{eqnarray}}
\def\eqa{\end{eqnarray}}

\newcommand{\bd}{\begin{displaymath}}
\newcommand{\ed}{\end{displaymath}}
\newcommand{\ba}{\begin{eqnarray}}
\newcommand{\ea}{\end{eqnarray}}

\def\ff{\widehat f}

\def\R{\mathbb{R}}

\newcommand{\setR}{\mathbb{R}}
\newcommand{\theq}{{\mathbf S}}

\begin{document}

\author{G.~Toscani}
\affiliation{Department of Mathematics, University of Pavia,  via Ferrata 1, Pavia, Italy. } \email[ e-mail address: ]{giuseppe.toscani@unipv.it}

\date{\today}

\title{Boltzmann legacy and wealth distribution}

\begin{abstract}
   We briefly review  results on nonlinear kinetic equation of Boltzmann type which describe the evolution of wealth in a simple agents market.
The mathematical structure of the underlying kinetic equations allows to use well-known techniques of wide use in kinetic theory of rarefied gases
to obtain information on the process of relaxation to a stationary profile, as well as to identify simple interaction rules which are responsible
of the formation of Pareto tails.
\end{abstract}

\maketitle

\section{Introduction}

In kinetic theory of rarefied gas, the evolution of the phase space density is described by the Boltzmann equation \cite{Cer, Cer94}
 $${\partial\over \partial t}f(t;x,v) = -v\cdot\nabla_x f(t;x,v) +
 Q (f(t;x,v)).$$
 This equation contains terms accounting for the two ways that the  density can change. The
 $$-v\cdot\nabla_x f(t;x,v)$$ term represents the effects of {\it  streaming};
 that is, the motion
 $$x_0 \mapsto x_0 + (t-t_0)v_0\qquad v_0 \mapsto v_0\eqno(1.2)$$
 of molecules between collisions. The $Q (f(x,v,t))$ term represents the effects
 of binary collisions and describes relaxation to the local Maxwellian equilibrium \cite{Cer, Cer94}.

The collision term $Q$ accounts for all kinematically possible (those that conserve both  momentum and energy) binary collisions. The
post--collisional velocities $v^*$ and $w^*$ are given by
 \be\label{co1}
 v^* = {1\over 2}(v+w+|v-w|{\bf n}) ,\qquad w^* = {1\over 2}(v+w-|v-w|{\bf n}),
 \ee
in which ${\bf n}$ is a unit vector.

For Maxwellian molecules \cite{Bobylev-Carrillo-Gamba, BMP02, BK00, BK02, BCT06}
 \be\label{max}
Q\bigl(f\bigr)(v) = \int_{{\R}^3\times{S^2}} B\left((v-w) \cdot {\bf n}\right)\bigl[ f( v^*)g( w^*)] \, d w d{\bf n} - \rho f(v).
 \ee
$B(\cdot)$ is a measure of the collision frequency,  $d{\bf n}$  denotes the {\it normalized} surface measure on the unit sphere $S^2$, and
$$\rho = \int_{\R^3}f(v)\, d v\ .$$
The Boltzmann description relies on a main \emph{ansatz}. While the number of gas molecules is enormously high so that collisions between
molecules do not retain memory, in view of rarefaction only binary collisions are important for the evolution.

Starting from the pioneering works of Mandelbrot \cite{mandel}, it is now commonly accepted by the kinetic community that in many aspects a
trading market composed of a sufficiently large number of agents can be described using the laws of statistical mechanics, just like a rarefied
gas, composed of many interacting particles. In fact, there is an almost literal translation of concepts: \emph{molecules} are identified with the
\emph{agents}, the \emph{particles'energy} correspond to the \emph{agents' wealth}, and binary \emph{collisions} translate into \emph{trade
interactions}. This modelling is clearly rather {\em ad hoc}, but if one is willing to accept the proposed analogies between trading agents and
colliding particles, then various well established methods from kinetic theory and statistical physics are ready for application to the field of
economy. Most notably, the numerous tools originally devised for the study of the energy distribution in a rarefied gas can now be used to analyze
wealth distributions. In this way, the kinetic description of market models via a Boltzmann-type equation provides one possible explanation for
the development of universal profiles in wealth distributions of real economies.

The distribution of wealth in a market economy is nowadays of great interest. Among other approaches, the description of market models via kinetic
equations is a fertile ground for research, well documented by numerous contributions in the recent books \cite{Basu,book1,book2,book3,book4}, or
by the introductory articles \cite{review, Gup06, hayes, encyclopedia}. The bridge between Maxwell molecules and nonconservative economies has
been outlined by F.$\null \,\,$Slanina \cite{slanina}. In this paper, a clear parallelism between the evolution of wealth in a simple economy and
the evolution of the particle density in a one-dimensional dissipative gas has been established. This paper motivated to eventually adapt more and
more of the ideas, which have been developed in the studies of dissipative Maxwell gases, to the economic framework.

It should be emphasized, however, that {\em there are\/} substantial differences between the collision mechanism for classical gas molecules and
the modelling of trade interactions. In the new framework, interactions can lack the usual microscopic conservation laws for (the analogues of)
momentum and energy; moreover, random effects play a crucial r\^ole. In fact, the key step in establishing a reasonable kinetic market model is
the definition of sensible rules on the {\em microscopic\/} level, i.e., the prescription of how wealth is exchanged in trades. Such rules are
usually derived from plausible assumptions in an {\em ad hoc\/} manner. (This is clearly in contrast to the original Boltzmann equation, where the
microscopic collisions are governed by the laws \fer{co1}.)

Nevertheless, strong analogies remain. Maybe the most important one is related to the fact that, as it happens in the classical Boltzmann equation
\cite{Cer, Cer94}, where relaxation to Maxwellian equilibrium is shown to be a universal behavior of the solution, here the corresponding output
of the model are the {\em macroscopic\/} statistics of the wealth distribution in the society, to which the solution is shown to relax. Mostly
important, while relaxation to equilibrium in the Boltzmann equation is achieved by looking at the monotonicity of the entropy functional,
relaxation to the steady wealth profile can be achieved by looking here at the monotonicity of new convex functionals.

 The comparison of this output with realistic data is up to now the only means to evaluate
--- {\em a posteriori} --- the quality of a proposed model. For instance, it is commonly accepted that the wealth distribution should approach a
stationary (or, in general, a self-similar) profile for large times, and that the latter should exhibit a {\em Pareto tail}. Such overpopulated
tails are a manifestation of the existence of an upper class of very rich agents, i.e.\ an indication of an unequal distribution of wealth. The
various articles in \cite{book1} provide an overview over historical and recent studies on the shape of wealth distributions; see also
\cite{review} for a collection of relevant references.

In general, the richness of the steady states for kinetic market models is another remarkable difference to the theory of Maxwell molecules
\cite{Bob}. While the Maxwell distribution is the universal steady profile for the velocity distribution of molecular gases, the stationary
profiles for wealth can be manifold, and are in general not explicitly known analytically. In fact, they depend heavily on the precise form of the
microscopic modelling of trade interactions. Consequently, in investigations of the large-time behavior of the wealth distribution, one is
typically limited to describe a few analytically accessible properties (e.g.\ moments and smoothness) of the latter.

A variety of models has been proposed and numerically studied in view of the relation between parameters in the microscopic rules and the
resulting macroscopic statistics. The features typically incorporated in kinetic trade models are saving effects and randomness. Saving means that
each agent is guaranteed to retain at least a certain minimal fraction of his initial wealth at the end of the trade. This concept has probably
first been introduced in \cite{CC00} (see also \cite{angle, DY00}), where a fixed saving rate for all agents has been proposed. Randomness means
that the amount of wealth changing hands is non-deterministic. Among others, this idea has been developed in \cite{CPT}, in order to include the
effects of a risky market. Depending upon the specific choice of the saving mechanism and the stochastic nature of the trades, the studied systems
produce wealth curves with the desired Pareto tail --- or not.

In this short review,  we describe the Boltzmann-like approach to wealth distribution,  and compare a selection of recently developed models. We
will mainly treat two different types of interactions. The first type is such that the binary trade is microscopically \emph{conservative}, while
the second type is such that the binary trade is \emph{conservative} in the statistical mean only.  In these situations, the mean wealth in the
model Boltzmann equation is preserved, and one expects the formation of a stationary profile.

In the class of pointwise conservative trades, we focus on the model designed by Chakraborti and Chakrabarti \cite{CC00}, and on variants of it.
In the class of conservative in the statistical mean, we focus on models with {\em risky investments}, originally introduced by Cordier, Pareschi
and the author \cite{CPT}. The applied analytical techniques, however, easily generalize to a broader class of conservative economic games. These
techniques have been applied in the current mathematical literature \cite{CPT,PT,MTa,MTb,DT,DMT, DMTb}, where kinetic econophysics has been
treated in the framework of Maxwell-type molecules.  The interest reader, who wishes to obtain a deeper understanding of the mathematical roots
(and possible extensions) of the applied tools, is referred e.g.\ to \cite{Vil, CarTos}.

The kinetic approach is complementary to the numerous theoretical and numerical studies that can be found in the recent physics literature on the
subject, from which it differs in several subtle points. In our point of view, the evolution of wealth dendity is entirely based on the spatially
homogeneous {\em Boltzmann equation\/} associated to the microscopic trade rules of the respective model. Agents on the market are treated as a
{\em continuum}, just like molecules in classical gas dynamics. Not only does this approach constitute the most natural generalization of the
classical ideas to econophysics. But moreover, it clarifies that certain peculiar observations made in ensembles of finitely many agents and in
numerical experiments (like the apparent creation of steady distributions of infinite average wealth, e.g.\ \cite{CCM,CCS,review}) are genuine
{\em finite size effects}.

For the sake of completeness, a comment on the justification of kinetic market models is in place. The socio-economic behavior of a (real)
population of agents is extremely complex. Apart from elements from mathematics and economics, a sound description --- if one at all exists ---
would necessarily need contributions from various other fields, including psychology. Clearly, the available mathematical models  are too simple
to even pretend to reflect the real situation. However, the idea to describe economic trades in terms of a kinetic equation gives rise to a
variety of challenging mathematical problems, both from the theoretical and numerical point of view. In particular, it is remarkable that this
class of simple models possesses such a wide spectrum of possible equilibria (some of which indeed resemble realistic wealth distributions).
Moreover, kinetic market models are extremely flexible with respect to the introduction of additional effects. In this way, the described models
should be considered as basic building blocks, that can easily be combined, adapted and improved.

\section{Boltzmann models for wealth}

We consider a class of models in which agents are indistinguishable. Then, an agent's ``state'' at any instant of time $t\geq0$ is completely
characterized by his current wealth $w\geq0$. When two agents encounter in a trade, their {\em pre-trade wealths\/} $v$, $w$ change into the {\em
post-trade wealths\/} $v^*$, $w^*$ according to the rule
\begin{equation}
  \label{eq.trules}
  v^* = p_1 v + q_1 w, \quad w^* = q_2 v + p_2 w.
\end{equation}
The {\em interaction coefficients\/} $p_i$ and $q_i$ are non-negative random variables. While $q_1$ denotes the fraction of the second agent's
wealth transferred to the first agent, the difference $p_1-q_2$ is the relative gain (or loss) of wealth of the first agent due to market risks.
We assume that $p_i$ and $q_i$ have fixed laws, which are independent of $v$ and $w$, and of time.

In one-dimensional models, the wealth distribution $f(t;w)$ of the ensemble coincides with agent density and satisfies the associated spatially
homogeneous Boltzmann equation of Maxwell type \fer{max}
\begin{equation}
  \label{eq.boltzmann}
 \partial_t f + f = Q_+(f,f),
\end{equation}
on the real half line, $w\geq0$. The collisional gain operator $Q_+(t;v)$, which quantifies the gain of wealth $v$ at time $t$ due to binary
trades, acts on test functions $\varphi(w)$ as
\begin{align}
\nonumber Q_+(f,f)[\varphi] :=&
  \int_{\setR_+} \varphi(w)Q_+\big(f,f\big)(w)\,dw\\
  \label{eq.qweak}
  =& \frac12 \int_{\setR_+^2} \langle \varphi(v^*) + \varphi(w^*) \rangle f(v)f(w)\,dv\,dw,
\end{align}
with $\langle\cdot\rangle$ denoting the expectation with respect to the random coefficients $p_i$ and $q_i$ in (\ref{eq.trules}). The large-time
behavior of the density is heavily dependent of the evolution of the average wealth
\begin{equation}
  \label{eq.conserve}
  M(t) := M_1(t) = \int_{\setR_+} w f(t;w)\,dw  ,
\end{equation}
Conservative models are such that the average wealth of the society is conserved with time, $M(t) = M$, where the value of $M$ is finite. In terms
of the interaction coefficients, this is equivalent to $\langle p_1+q_2\rangle = \langle p_2+q_1\rangle = 1$ .

The Boltzmann equation \fer{eq.boltzmann} belongs to the Maxwell type. As briefly described in the Introduction, in the Boltzmann equation for
Maxwell molecules the collision frequency is independent of the relative velocity \cite{Bob}, and the loss term in the collision operator is
linear. This introduces a great simplification, that allows to use most of the well established techniques developed for the three-dimensional
spatially homogeneous Boltzmann equation for Maxwell molecules in the field of wealth redistribution.

\subsection{Pointwise conservative models}
The first explicit description of a binary wealth exchange model dates back to Angle \cite{angle} (although the intimate relation to statistical
mechanics was only described about one decade later \cite{IKR,DY00}): in each binary interaction, winner and loser are randomly chosen, and the
loser yields a random fraction of his wealth to the winner. From here, Chakraborti and Chakrabarti \cite{CC00} developed the class of {\em
strictly conservative\/} exchange models, which preserve the total wealth in each individual trade,
\begin{equation}
  \label{eq.strict}
  v^* + w^* = v + w .
\end{equation}
In its most basic version, the microscopic interaction is determined by one single parameter $\lambda\in(0,1)$, which is the global {\em saving
propensity}. In interactions, each agent keeps the corresponding fraction of his pre-trade wealth, while the rest $(1-\lambda)(v+w)$ is equally
shared among the two trade partners,
\begin{equation}
  \label{eq.xchg}
  v^* = \lambda v + \frac12(1-\lambda)(v+w), \quad   w^* = \lambda w + \frac12(1-\lambda)(v+w) .
\end{equation}
In result, all agents become equally rich eventually. Non-deterministic variants of the model have been proposed, where the amount
$(1-\lambda)(v+w)$ is not equally shared, but in a stochastic way:
\begin{equation}
  \label{eq.strictxchg}
  v^* = \lambda v + \epsilon(1-\lambda)(v+w), \quad w^* = \lambda w + (1-\epsilon)(1-\lambda)(v+w),
\end{equation}
with a random variable $\epsilon\in(0,1)$.

\subsection{Conservative in the mean models}

Cordier et al.\ \cite{CPT} have introduced the CPT model, which breaks with the paradigm of strict conservation. The idea is that wealth changes
hands for a specific reason: one agent intends to {\em invest\/}\nobreakspace his wealth in some asset, property etc.\ in possession of his trade
partner. Typically, such investments bear some risk, and either provide the buyer with some additional wealth, or lead to the loss of wealth in a
non-deterministic way. An easy realization of this idea \cite{MTa} consists in coupling the previously discussed rules (\ref{eq.xchg}) with some
{\em risky investment\/} that yields an immediate gain or loss proportional to the current wealth of the investing agent,
\begin{equation}
  \label{eq.cpt}
  v^* = \Bigl(\lambda +\eta_1\Bigr)v + (1-\lambda)w, \quad
  w^* = \Bigl(\lambda +\eta_2\Bigr)w + (1-\lambda)v,
\end{equation}
The coefficients $\eta_1,\eta_2$ are random parameters, which are independent of $v$ and $w$, and distributed so that always $v^*,\,w^*\geq0$,
i.e.\ $\eta_1,\,\eta_2\geq -\lambda$. For centered $\eta_i$,
\begin{equation}\label{cons1}
  \langle v^* + w^* \rangle = (1+\langle\eta_1\rangle) v + (1+\langle\eta_2\rangle) w = v + w ,
\end{equation}
implying conservation of the average wealth. Various specific choices for the $\eta_i$ have been discussed \cite{MTa}. The easiest one leading to
interesting results is $\eta_i=\pm\mu$, where each sign comes with probability $1/2$. The factor $\mu\in(0,\lambda)$ should be understood as the
{\em intrinsic risk\/} of the market: it quantifies the fraction of wealth agents are willing to gamble on.

\

\section{Boltzmann equilibria}
In conservative markets, where $M(t) = M$, the details of the binary trade determine the profile of the steady wealth distribution. We introduce
the characteristic function
\begin{equation}
  \label{eq.q}
  \theq(s) = \frac12 \Big( \sum_{i=1}^2 \langle p_i^s+q_i^s \rangle \Big) - 1 ,
\end{equation}
which is convex in $s>0$, with $\theq(0) = 1$. Also, $\theq(1) = 0$ because of the conservation property \eqref{cons1}. The results from
\cite{MTa,DMT} imply the following. Unless $\theq(s)\geq0$ for all $s>0$, any solution $f(t;w)$ tends to a steady wealth distribution
$P_\infty(w)=f_\infty(w)$, which depends on the initial wealth distribution only through the conserved mean wealth $M>0$. Moreover, exactly one of
the following is true:
\begin{itemize}
\item[(PT)] if $\theq(\alpha)=0$ for some $\alpha>1$,
  then $P_\infty(w)$ has a {\em Pareto tail\/} of index $\alpha$;
\item[(ST)] if $\theq(s)<0$ for all $s>1$,
  then $P_\infty(w)$ has a {\em slim tail};
\item[(DD)] if $\theq(\alpha)=0$ for some $0<\alpha<1$,
  then $P_\infty(w)=\delta_0(w)$, a {\em Dirac Delta\/} at $w=0$.
\end{itemize}

In case (PT), exactly the moments $M_s(t)$ with $s>\alpha$ blow up as $t\to\infty$, giving rise to a Pareto tail of index $\alpha$. We emphasize
that $f(t;w)$ possesses finite moments of all orders at any finite time. The Pareto tail forms {\em in the limit $t\to\infty$}.

In case (ST), all moments converge to limits $M_s(t)\to M_s^*$, so the tail is slim.

In case (DD), all moments $M_s(t)$ with $s>1$ blow up. The underlying process is a separation of wealth as time increases: while more and more
agents become extremely poor, fewer and fewer agents possess essentially the entire wealth of the society.

\subsection{The legacy of Maxwell molecules}

The main tool to obtain the aforementioned results is the use of the Fourier transform. This idea, which goes back to the seminal work of Bobylev
\cite{Bo75, Bob}, is well-suited to treat collision kernels of Maxwellian type. In particular, the Fourier representation is particularly adapted
to the use of various Fourier metrics.

According to the collision rule \eqref{eq.trules}, the transformed gain term reads
\begin{equation}
  \label{trascoll}
  \widehat{Q}_+\bigl( \ff,\ff \bigr)(\xi) = \frac 12 \left\langle
    \widehat{f}(p_1\xi)
    \widehat{f}(q_1\xi)  +  \widehat{f}(p_2\xi) \widehat{f}(q_2\xi)\right\rangle  .
\end{equation}
The initial conditions turn into
\[
\widehat{f}_0(0)=1\quad\mbox{and}\quad \widehat{f_0}^\prime(0) = iM .
\]
Hence, the Boltzmann equation  \fer{eq.boltzmann} can be rewritten as
\begin{equation}
  \label{fkac}
  \frac{\partial \widehat f(t;\xi)}{\partial t}  + \widehat f(t;\xi) = \frac 12 \big\langle
    \widehat{f}(p_1\xi)
    \widehat{f}(q_1\xi)  +  \widehat{f}(p_2\xi) \widehat{f}(q_2\xi)\big\rangle.
\end{equation}

Equation \fer{fkac} can be easily treated from a mathematical point of view owing to the well-known techniques introduced so far to study the
Boltzmann equation for Maxwell molecules and its caricatures, mainly Kac equation \cite{Kac}.

\medskip
{\bf Acknowledgement.} This work has been done under the activities of the National Group of Mathematical Physics (GNFM). The support of the MIUR
project ``Variational, functional-analytic, and optimal transport methods for dissipative evolutions and stability problems'' is kindly
acknowledged.

\end{document}